\documentclass[11pt, a4paper, twocolumn]{article}
\usepackage[utf8]{inputenc}
\usepackage{hyperref}
\usepackage{amsmath}
\usepackage{physics}
\usepackage{amsfonts}
\usepackage{multicol}
\usepackage{multirow}
\usepackage{graphicx}
\usepackage{abstract}
\usepackage{mathtools}
\usepackage{algorithm}
\usepackage{booktabs}
\usepackage{algpseudocode}
\usepackage[dvipsnames,svgnames]{xcolor}
\usepackage[sorting=none, style=numeric-comp]{biblatex}
\bibliography{refs}
\usepackage[affil-it]{authblk} 

\hypersetup{pdftitle={Quantum Diffusion Models}}

\title{Quantum Diffusion Models}

\author[1]{Andrea Cacioppo\thanks{}}
\author[1]{Lorenzo Colantonio}
\author[1,2,3]{Simone Bordoni}
\author[1,3]{Stefano Giagu}

\affil[1]{\small Department of Physics, Sapienza Universit\`a di Roma, Roma, Italy}
\affil[2]{\small QRC, Technology Innovation Institute, Abu Dhabi, UAE}
\affil[3]{\small INFN Sezione di Roma, Roma, Italy}
\date{}

\begin{document}

\twocolumn[
\maketitle

\begin{@twocolumnfalse}
\begin{abstract}
We propose a quantum version of a generative diffusion model. In this algorithm, artificial neural networks are replaced with parameterized quantum circuits, in order to directly generate quantum states. We present both a full quantum and a latent quantum version of the algorithm; we also present a conditioned version of these models. The models' performances have been evaluated using quantitative metrics complemented by qualitative assessments. An implementation of a simplified version of the algorithm has been executed on real NISQ quantum hardware.\\ \vspace{10mm}
\end{abstract}
\end{@twocolumnfalse}]

\section{Introduction}
\let\thefootnote\relax\footnotetext{*Email: \textit{andrea.cacioppo@uniroma1.it}}
The ability to reproduce distributions from a dataset and to sample from them is one of the most useful tasks of machine learning (ML) models; these are referred to as generative models. In this context, diffusion models are algorithms inspired by nonequilibrium thermodynamics \cite{sohl2015deep, ho2020denoising, 10081412, song2022denoising} that offer a promising alternative to the established variational autoencoders (VAEs) \cite{kingma2013auto} and generative adversarial networks (GANs) \cite{goodfellow2020generative}, due to their superior sample quality and variability. \\
Parallel to these advancements in ML, the field of quantum computing has experienced a significant interest. Despite its rapid development in recent years, it is still early in its exploration, with much to learn about its possibilities and limits. Key milestones such as quantum error correction \cite{qec2, qec1, qec3, qec4, qec5} have yet to be achieved, underlining the emergent nature of this technology.
The current generation of quantum processors (NISQ)\cite{preskill2018quantum} is too noisy for the implementation of quantum supremacy algorithms \cite{Shor_1997}. On the other hand, quantum machine learning (QML) algorithms are less sensitive to noisy devices \cite{Biamonte_2017}. Even if these algorithms have not proven to outperform their classical counterparts, many interesting results have been obtained both in simulations and on NISQ devices \cite{qml0, qml1, qml2, qml3, qml4, qml5, qml6, qml7}. The first meaningful utilization of quantum computing may likely come from algorithms of this kind, complemented by error mitigation techniques. \cite{em1, em2, em3, em4}.\\
The quantum versions of some established generative algorithms have already been developed, like the quantum GAN \cite{dallaire2018quantum, Bravo_Prieto_2022} and the quantum VAE \cite{khoshaman2018quantum}, however, initial attempts to develop a quantum version of a diffusion model are limited to basic or simplified scenarios \cite{parigi2023quantum, zhang2023generative}. In this work we present a quantum diffusion model (QDM), which can be employed for the generation of classical data or quantum data directly, in two different versions. We show the application of our approach for generating samples from the MNIST dataset and demonstrate a simplified version of this method, implemented on real quantum hardware.\\

This work is organized as follows:\\
in section \ref{sec:background} we provide a high level explanation of classical diffusion models and parameterized quantum circuits.\\
In section \ref{sec:QDM} we describe our QDM and the main differences with the classical model.\\
In section \ref{sec:simulation} we present the results obtained with the QDM in simulation, for different configurations, along with the metrics we used to evaluate each model.\\
In section \ref{sec:hardware} we describe the algorithm adaptations necessary to execute it on a NISQ device and discuss the results obtained.

\section{Background}\label{sec:background}

In this section, we review the main aspects of classical diffusion models and parameterized quantum circuits, necessary to understand the proposed quantum diffusion model.

\subsection{Parameterized quantum circuits}\label{sec:PQC}
\begin{figure}[htb]
\includegraphics[width=0.4\textwidth]{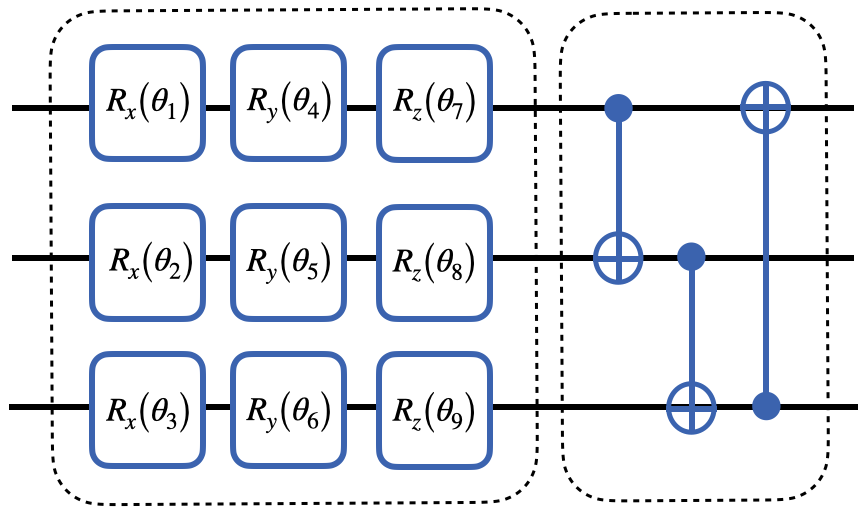}
\centering
\caption{Example of a PQC layer ansatz for three qubits. The rotation gates contain classical, trainable, parameters while the C-NOT gates create entanglement between the qubits.}
\label{fig:ansatz}
\end{figure}

A parameterized quantum circuit (PQC) \cite{preskill2018quantum}\cite{benedetti2019parameterized} is a trainable algorithm that implements a parametric transformation on a quantum state, which is the result of the application of quantum gates, generally organized in layers. For this reason it can be considered as the quantum counterpart of an artificial neural network (ANN). 
In our work we have tested different layer ansatzes, obtaining the best results with the strongly entangling ansatz \cite{schuld2020circuit}, composed of trainable rotation gates ($\hat{R}_x$, $\hat{R}_y$, $\hat{R}_z$), acting on all qubits, followed by a series of C-NOT gates, coupling neighboring qubits, as represented in figure \ref{fig:ansatz}.
\\
In principle, it would be possible to train a PQC directly on quantum hardware using techniques like the parameter-shift rule \cite{mitarai2018quantum, schuld2019evaluating}. However, computing gradients on current quantum hardware is unfeasible because of the high level of noise. For this reason, in our work, we trained the PQCs in simulations using the software library \textit{PennyLane} \cite{bergholm2018pennylane}. Both the forward and the optimization steps are performed on a classical computer with techniques based on gradient descent. The trained PQCs can then be employed for sampling both in simulation and on quantum hardware.\\

When a PQC is applied to process classical data, it is necessary to add a state preparation and a measurement part, to respectively encode and decode classical information into and from quantum states. Throughout this work, we have used amplitude encoding \cite{schuld2021machine} to encode classical information into quantum states. This consists in writing a classical vector's components as the coefficients of a quantum state.
\begin{equation*}
    \textbf{x} \longrightarrow \lvert \textbf{x}\rangle = \sum_{i=1}^{2^N} x_i \lvert \textbf{i} \rangle
\end{equation*}
where $N$ is the number of qubits and $\lvert \textbf{i} \rangle$ represents the ordered vector of the computational basis. This way it is possible to encode $2^N$ classical features into $N$ qubits. However, to perform this encoding, it is necessary to use a number of C-NOT gates that grows exponentially in the number of qubits \cite{PhysRevA.103.032430}.

\subsection{Classical diffusion model}\label{sec:CDM}
A diffusion model \cite{ho2020denoising} is a generative model \cite{ruthotto2021introduction}, namely an algorithm that is used to learn the probability distribution $p(\textbf{x})$ associated to a dataset, with the objective of sampling from this distribution.\\
The main idea of this method is to consider a diffusion process, represented by a Markov chain, which maps an arbitrary distribution $q(\textbf{x}_0)$ to a treatable distribution $\pi(\textbf{x}_T)$, e.g. a Gaussian multivariate distribution. This is achieved through a Markov kernel $q(\textbf{x}_{t}|\textbf{x}_{t-1})$, with $t \in \{1, \dddot{} T\}$.
A parametric model is then trained to reproduce the inverse Markov chain (reverse trajectory), estimating the inverse transition probability $p_{\boldsymbol{\theta}}(\textbf{x}_{t-1}|\textbf{x}_t)$. \\
Knowing $p_{\boldsymbol{\theta}}(\boldsymbol{x}_{t-1}|\boldsymbol{x}_t)$, the target distribution $p_{\boldsymbol{\theta}}(\textbf{x}_0)$ can be computed as:
\begin{equation*}
    p_{\boldsymbol{\theta}}(\textbf{x}_0) = \int d\textbf{x}_{1:T}\hspace{0.1cm}\boldsymbol{\pi}(\textbf{x}_T)\prod_{t=1}^Tp_{\boldsymbol{\theta}}(\boldsymbol{x}_{t-1}|\boldsymbol{x}_t)
\end{equation*}
A common choice for the forward conditional probability is the Gaussian kernel:
\begin{equation}\label{eq:kernel}
q(\textbf{x}_{t}|\textbf{x}_{t-1}) = \mathcal{N}(\mathbf{x}_{t}; \sqrt{1-\beta_t}\mathbf{x}_{t-1}, \beta_t\mathbf{I})
\end{equation}
where $\beta_1, \dddot{}, \beta_T$ is a variance schedule.
This choice allows to sample $\mathbf{x}_t$ at an arbitrary timestep $t$ in closed form:
\begin{equation}\label{eq:closed}
\begin{cases}
\mathbf{x}_t(\mathbf{x}_0, \epsilon) = \sqrt{\bar{\alpha}_t}\mathbf{x}_0 + \sqrt{1 - \bar{\alpha}_t} \boldsymbol{\epsilon}\\
\boldsymbol{\epsilon} \sim \mathcal{N}(\mathbf{0}, \mathbf{1})
\end{cases}
\end{equation}
with $\bar{\alpha}_t \equiv \prod_{s=1}^t(1-\beta_s)$.
When $\beta_t$ is sufficiently small, the inverse conditional probability has the same functional form \cite{sohldickstein2015deep} of the forward kernel, hence:
\begin{equation*}
    p_{\boldsymbol{\theta}}(\textbf{x}_{t-1}|\textbf{x}_t) = \mathcal{N}(\mathbf{x}_{t-1}; \boldsymbol{\mu}_{\boldsymbol{\theta}}(\mathbf{x}_t, t), \sigma^2_t\mathbf{I} )
\end{equation*}
where the mean can be further parameterized as:
\begin{equation}\label{eq:mu}
    \boldsymbol{\mu}_{\boldsymbol{\theta}}(\mathbf{x}_t, t) = \frac{1}{\sqrt{\alpha_t}}\Big(\mathbf{x}_t-\frac{1-\alpha_t}{\sqrt{1-\bar{\alpha}_t}}\boldsymbol{\epsilon}_{\boldsymbol{\theta}}(\mathbf{x}_t, t)\Big)
\end{equation}
with $\alpha_t \equiv 1-\beta_t$. In this formula $\boldsymbol{\epsilon}_{\boldsymbol{\theta}}(\mathbf{x}_t, t)$ is a function approximator, implemented by an ANN, intended to predict $\boldsymbol{\epsilon}$ from $\mathbf{x}_t$, with $\boldsymbol{\epsilon} \sim \mathcal{N}(\mathbf{0}, \mathbf{I})$. This is equivalent to minimizing the variational bound on negative log-likelihood:
\begin{align*}
    \mathbb{E}\big[-&\text{log}\hspace{0.1cm}p_{\boldsymbol{\theta}}(\textbf{x}_0)\big]\leq \\
    \leq\mathbb{E}_q\Bigg[-\text{log}\hspace{0.1cm}p(\textbf{x}_T)&-\sum_{t\geq 1}\text{log}\hspace{0.1cm}\frac{p_{\boldsymbol{\theta}}(\boldsymbol{x}_{t-1}|\boldsymbol{x}_t)}{q(\textbf{x}_{t}|\textbf{x}_{t-1})}\Bigg]\equiv L\nonumber 
\end{align*}
After training the model, one can sample from $p_{\boldsymbol{\theta}}(\textbf{x}_0)$ by extracting a data point from the analytically treatable distribution $\textbf{x}_T \sim \pi(\textbf{x}_T)$ and applying the inverse Markov chain, as described in Algorithm \ref{alg:sample}.

\begin{algorithm}
\caption{Sampling algorithm}\label{alg:sample}
\begin{algorithmic}
\State{Sample $\mathbf{x}_{T}$ $\sim$ $\pi(\mathbf{x}_{T})$ }
\For{$t=T,...,1$}
\State{Sample $\mathbf{z} \sim \mathcal{N}(\mathbf{0},\mathbf{I})$ if $t>1$, else $\mathbf{z}=0$}
\\
\State $\mathbf{x}_{t-1}=\frac{1}{\sqrt{\alpha_t}}\big(\mathbf{x}_{t}-\frac{1-\alpha_t}{\sqrt{1-\overline{\alpha}_t}}\epsilon_{\theta}(\mathbf{x}_{t},t)\big) +\sigma_t \mathbf{z}$
\EndFor
\end{algorithmic}
\end{algorithm}

\section{Quantum diffusion model}\label{sec:QDM}
The quantum adaptation of a diffusion model works through direct manipulation of quantum states. This is accomplished by substituting the traditional ANN with a parameterized quantum circuit. Due to the intrinsic differences between these two models, several adjustments are necessary to obtain a working algorithm.

\subsection{Training}

In a classical diffusion model, as discussed in section \ref{sec:CDM}, a forward Markov chain is constructed by adding noise through a Markov kernel (Eq. \ref{eq:kernel}). 
At the same time, an ANN is trained to approximate the Gaussian kernel mean (Eq. \ref{eq:mu}).
This procedure is not possible in a quantum model, as it would involve decoding and encoding quantum states, at each time step, to perform classical operations on the vectors. This is impractical as current NISQ devices are subject to measurement errors and the encoding of classical data is inefficient \cite{cortese2018}. For this reason, we have decided to utilize a hybrid approach, implementing the forward Markov chain classically, while utilizing a PQC in the backward process and the sampling phase. The PQC defines an operator $\hat{P}(\boldsymbol{\theta}, t)$, that performs data denoising directly:
\begin{equation*}
    \hat{P}(\boldsymbol{\theta}, t)\lvert \textbf{x}_t\rangle = \lvert \textbf{x}_{t-1}\rangle
\end{equation*}
Here $\ket{\textbf{x}_t}$ is the quantum state obtained by encoding the classical vector $\textbf{x}_t$, which represents the $t$ step in the forward chain.
This way, the complete backward process becomes a denoising process, which can be performed by the consecutive application of $T$ different PQCs. 
\begin{equation*}\label{eq:denoising}
    \hat{P}(\boldsymbol{\theta}, t=1)\dddot{} \hat{P}(\boldsymbol{\theta}, t=T)\lvert \textbf{x}_T\rangle = \lvert \textbf{x}_0\rangle
\end{equation*}
We observed that training a single circuit for all denoising steps results in only a limited decline in performance.
Another important consideration is that, in the iterative application of the PQC, the coefficients of the quantum states can become complex, even if the initial state $\mathbf{x}_0$ had real values. For this reason we have tested an alternative to the classical forward kernel, as described in equation \ref{eq:closed}, where we used complex noise, instead of real noise, at every time step of the training:
\begin{equation*}
\begin{cases}
\boldsymbol{\epsilon} = \boldsymbol{\epsilon}_r+ i\boldsymbol{\epsilon}_i\\
\boldsymbol{\epsilon}_r, \boldsymbol{\epsilon}_i\sim \mathcal{N}(\mathbf{0}, \mathbf{1})
\end{cases}
\end{equation*}
 this approach has led to a significant improvement in performance.\\

The PQC has been trained using a completely classical optimizer. However, during the algorithm’s design, we have chosen a loss function that could be easily computed using a full quantum approach, namely the infidelity loss:
$$
L(\boldsymbol{\theta}) = 1-\mathbb{E} \big[F\big(\hat{P}(\boldsymbol{\theta}, t) \lvert\textbf{x}_{t+1}\rangle, \lvert\textbf{x}_t\rangle\big)\big]
$$
where $F$ indicate the quantum fidelity and the expected value is taken over all possible choices of $\mathbf{x}_t$ at every value of $t$ and $\mathbf{x}_0$.
On quantum hardware, the fidelity can be computed efficiently with the use of a multi-qubit swap test \cite{barenco1997stabilization}.

\subsection{Quantum circuit ansatz}\label{sec:ansatz}
\begin{figure*}[t!]
     \centering
     \includegraphics[width=\textwidth]{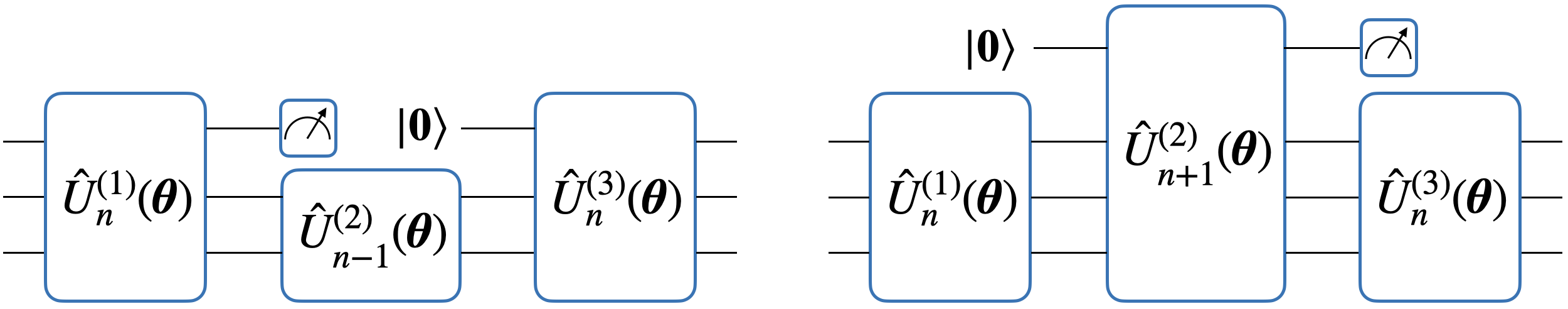}
     \caption{Representation of the bottleneck (left) and reverse-bottleneck architectures (right) using $m=1$. Each unitary transformation block in the image employs the layered structure reported in Figure \ref{fig:ansatz}.}
     \label{fig:PQC}
\end{figure*}
We have tested several architectures for the PQCs, the two circuits leading to the best performance are represented in figure \ref{fig:PQC}.
Both architectures use three unitary blocks operating on different qubits. In the first ansatz (bottleneck PQC), after a first unitary block $\hat{U}^{(1)}_n(\boldsymbol{\theta})$ acting on all qubits, the first $m$ qubits are measured, reducing the number of qubits to $n-m$. After the application of the second unitary block $\hat{U}^{(2)}_{n-m}(\boldsymbol{\theta})$, acting as a bottleneck, $m$ ancillary qubits are introduced in the state $|0\rangle^{\otimes m}$, and a final block $\hat{U}^{(3)}_n(\boldsymbol{\theta})$ is applied to all $n$ qubits. 
In the second ansatz (reverse-bottleneck PQC), the order of the ancillary qubit introduction and the measurements is reversed. This way, the central block $\hat{U}^{(2)}_{n+m}(\boldsymbol{\theta})$ acts on $n+m$ qubits. We observed little difference in the performance of the two architectures, reverse-bottleneck PQC performing slightly better. We have found that the best value for the number of measured qubits is $m = 1$. For these reasons, every result in the following refers to the reverse-bottleneck architecture, using one measured qubit.

Testing different PQC architectures, we have observed that the number of intermediate measurements of ancillary qubits have a strong impact on the performance of the algorithm. The introduction of intermediate measurements improves significantly the quality of generated samples. This can be explained by the fact that the measurement process acts as a non-linear map over the states amplitudes \cite{schuld2021machine}. Our numerical simulations have confirmed the non-linearity of these maps when using our ansatzes with trained parameters. Another remarkable finding is that measurements reduce the samples variability. In particular, when their overall number exceeds a certain threshold (dependent on the architecture and number of qubits) no sensible variability is observed in the generated samples, resulting in a model collapse. One potential explanation for this phenomenon is that, every time a qubit is measured, it is reset to a fixed state. This process erases part of the information contained in the initial noise \cite{nielsen1996entanglement}. Consequently, by increasing the number of measurements, the initial information is gradually lost, resulting in a loss of variance in the output states.
\\
Measurements of the ancillary qubits can be performed with two methods, namely with or without branch selection \cite{schuld2021machine}. With branch selection, we indicate the process of measuring the ancillary qubits and accepting the remainder of the state only if the outcome coincides with a predetermined choice. If the state is rejected, the circuit is executed again from the beginning. By measurement without branch selection, we indicate the process of accepting the remainder of the quantum state regardless of the measurement outcome. 
We have empirically observed that utilizing measurements with branch selection leads to better results. However, when performing branch selection on quantum hardware, each measurement has an average success rate $0.5$. Hence this process is not suitable for deep circuits, as the probability of accepting the circuit execution diminishes exponentially with the number of measurements. For this reason, throughout this work, we have used measurements without branch selection for the ancillary qubits. 
Another important remark is that, when performing measurements without branch selection, the output state is nondeterministic. This characteristic aligns our model more closely with the classical diffusion model.

\subsection{Model variations}
In this section, we present two independent variations on the algorithm described so far: a latent classical-quantum version and a conditioned version.

\paragraph{Latent models} To increase the expressive power of the QDM, it is possible to use a hybrid approach, by employing a pre-trained classical autoencoder \cite{schmidhuber2015deep} (latent model). With this approach, the QDM is trained on a low dimensional representation of data. The lower dimensionality of the problem, along with the strong non-linearity introduced by the classical autoencoder, has improved the quality of samples and allowed the use of smaller PQCs. Moreover, the consequent simplification of the PQC, has allowed the implementation of the algorithm on real quantum hardware, as explained in section \ref{sec:hardware}.

\paragraph{Conditioned models}\label{sec:conditioned}
Both the full quantum model and the latent model can be easily conditioned by increasing the dimension of the Hilbert space. This is done by taking the tensor product between the input quantum state and an additional state that encodes the label (label qubits):

\begin{equation}
\lvert \textbf{x}_t\rangle \longrightarrow \lvert k\rangle \otimes \lvert \textbf{x}_t\rangle
\end{equation}

where $\lvert k\rangle$ is the $k^{th}$ state of the computational basis. In order to encode $N$ labels, it is necessary to add $\lceil log_2(N) \rceil$ qubits. We emphasize that, as for the ancillary qubits, it is possible to measure label qubits with or without branch selection. We have chosen to use measurements without branch selection in our hardware implementation, and measurements with branch selection in our simulations.

\subsection{Model evaluation}\label{sec:evaluation}
To assess the performance of the proposed models, we utilized a combination of qualitative evaluation and quantitative metrics. To visually compare the distributions of the original and generated samples and enhance our evaluation of image quality, we employ dimensionality reduction techniques such as PCA \cite{jolliffe2016principal} and t-SNE \cite{van2008visualizing}.\\
On the other hand, for the quantitative assesment, we have employed the ROC-AUC \cite{bradley1997use}, to evaluate the conditioning performance, while for the generation performance, we have used both the Fréchet Inception Distance (FID) \cite{heusel2017gans} and the 2-Wasserstein distance for Gaussian mixture models (WaM) \cite{luzi2023evaluating}.  
To evaluate conditioned models, the ROC-AUC is computed using 10 binary classifiers, each specialized in distinguishing a single digit within the MNIST dataset. In this context, we assume the classifiers to be perfect, with each achieving an accuracy well above $0.99$ on the original dataset. This assumption enables the trained classifiers to serve as a reliable ground truth against which the conditioned samples can be evaluated.
To evaluate the generation performance we have employed both the FID and the WaM. We have chosen this second metric as the FID suffers main problems, as pointed out in \cite{heusel2017gans}, in particular with small and greyscale images. For the calculation of both metrics for the conditioned and unconditioned latent model, we employed a total of 70,000 samples, 10,000 at a time to evaluate satistics. For the fully quantum model we employed a total of 10,000 samples, 1000 at a time.
\begin{table*}[tp]
\centering
\caption{PQC characteristics of different models. The number of layers are referred to the architectures represented in figure \ref{fig:PQC}.}
\label{tab:params}
\begin{tabular}{cccc}
\toprule
\textbf{Feature} & \textbf{Full quantum} & \textbf{Unconditioned latent}  & \textbf{Conditioned latent} \\
\midrule
Layers & 30+40+30 & 15+20+15 & 15+20+15\\
Qubits & 8 & 3 & 3+4 \\
Parameters & 2520 & 510 & 1110 \\
Denoising steps (T) & 15 & 8 & 8 \\
Images size & 16x16 & 28x28 & 28x28 \\
Latent space & N.A. & 8 & 8 \\
Digits & ``0"/``1" & All & All\\
\bottomrule
\end{tabular}
\end{table*}

\section{Simulation Results}\label{sec:simulation}
In this section, we present the results obtained in simulation for the full quantum model without conditioning, and the latent model with and without conditioning. We omit the results obtained with the conditioned full quantum model, as adding a conditioning on only two labels did not give relevant results differences.
We have chosen to train each model on the MNIST dataset. For the full quantum model, the samples have been downsized from 28x28 pixels to 16x16 pixels to be encoded into 8 qubits using amplitude encoding. Moreover, only zeros and ones have been used to further simplify the problem.
For the latent models, we have chosen to use latent vectors of dimension 8, encoded using 3 qubits. In this case, we used every digit of the MNIST dataset. 
The conditioned version of the latent model required $4$ additional qubits to encode the labels ($10$ digits), for a total of $7$ qubits.
We have tested different number of layers for the unitary blocks of the reverse-bottleneck circuit ansatz (section \ref{sec:ansatz}). For the latent models, as the PQC acts on a smaller Hilbert space, it is possible to obtain optimal performances with a total number of $50$ layers. The full quantum model requires a total number of $100$ layers in order to reach sufficient expressive power. Even if these circuits are composed of a high number of layers, we have not experienced any issue related to barren plateaus \cite{mcclean2018barren} during training.
A summary of the characteristics of the models are reported in table \ref{tab:params}.\\
To convert quantum states into classical vectors, we have adopted the convention of considering the absolute values of the quantum state's amplitudes. Experimentally, this is equivalent to taking the square roots of the measurement frequencies in the computational basis.

\subsection{Full quantum model}\label{sec:full_quantum}

\begin{figure}[htb]
\includegraphics[width=0.45\textwidth]{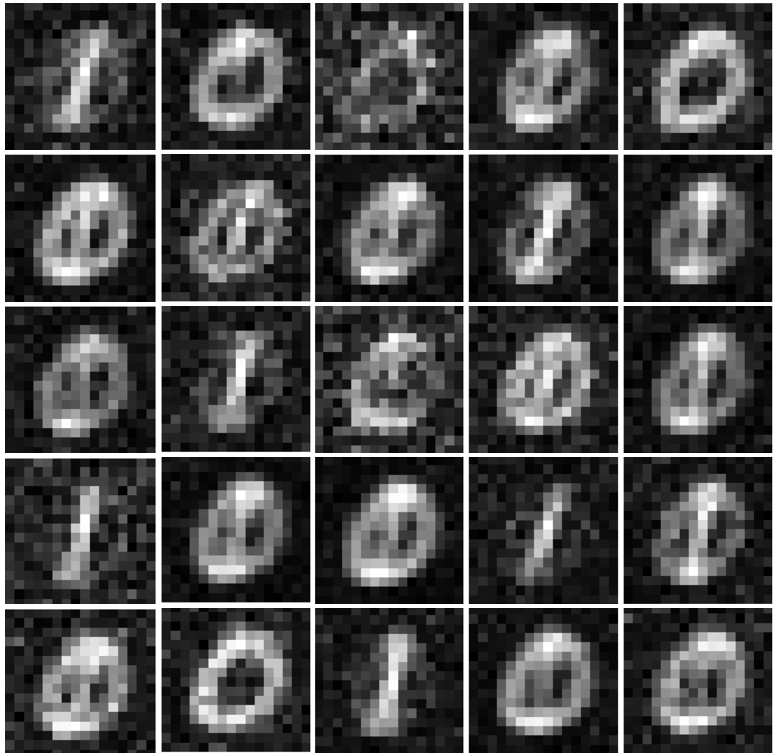}
\centering
\caption{Samples generated from the full quantum model.}
\label{fig:base_samples}
\end{figure}

\begin{figure*}[ht]
\includegraphics[width=0.96\textwidth]{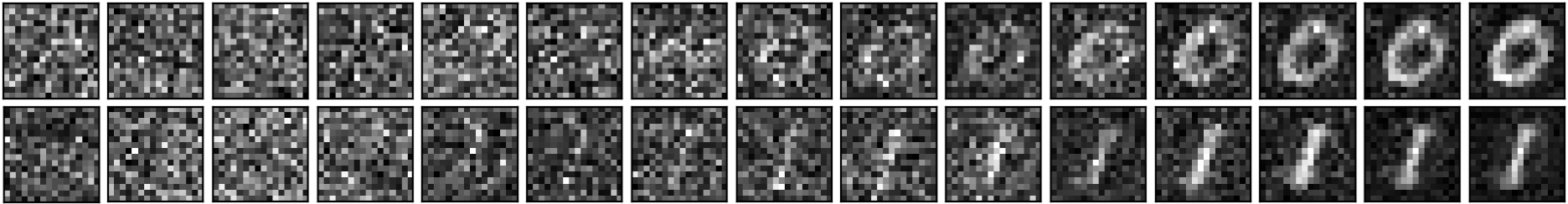}
\centering
\caption{Detail on the denoising process of a ``zero" and a ``one" digit for the full quantum model.}
\label{fig:denoising}
\end{figure*}

Samples generated by the full quantum model are shown in figure \ref{fig:base_samples}. The characteristics of the model used for the generation of these samples are reported in the first column of table \ref{tab:params}. We highlight how the number of time steps $T$ is significantly smaller with respect to classical diffusion models.
Two examples of the denoising process are represented in figure \ref{fig:denoising}.\\
The value of the FID between the original and generated datasets can be found in the first column of table \ref{tab:metrics}. 

\begin{table*}[htp]
\centering
\caption{Evaluation metrics for different models}
\label{tab:metrics}
\begin{tabular}{cccc}
\toprule
&\textbf{Full quantum} & \textbf{Unconditioned latent} & \textbf{Conditioned latent} \\
\midrule
FID & $256.6\pm2.0$ & $41.4\pm0.3$  & $38.2\pm 2.7$ \\
WaM & $240.3\pm2.4$ & $25.8\pm0.8$ & $13.5\pm1.1$ \\
\bottomrule
\end{tabular}\label{tab:metrics}
\end{table*}


\subsection{Unconditioned latent model}

\begin{figure}[htb]
\includegraphics[width=0.45\textwidth]{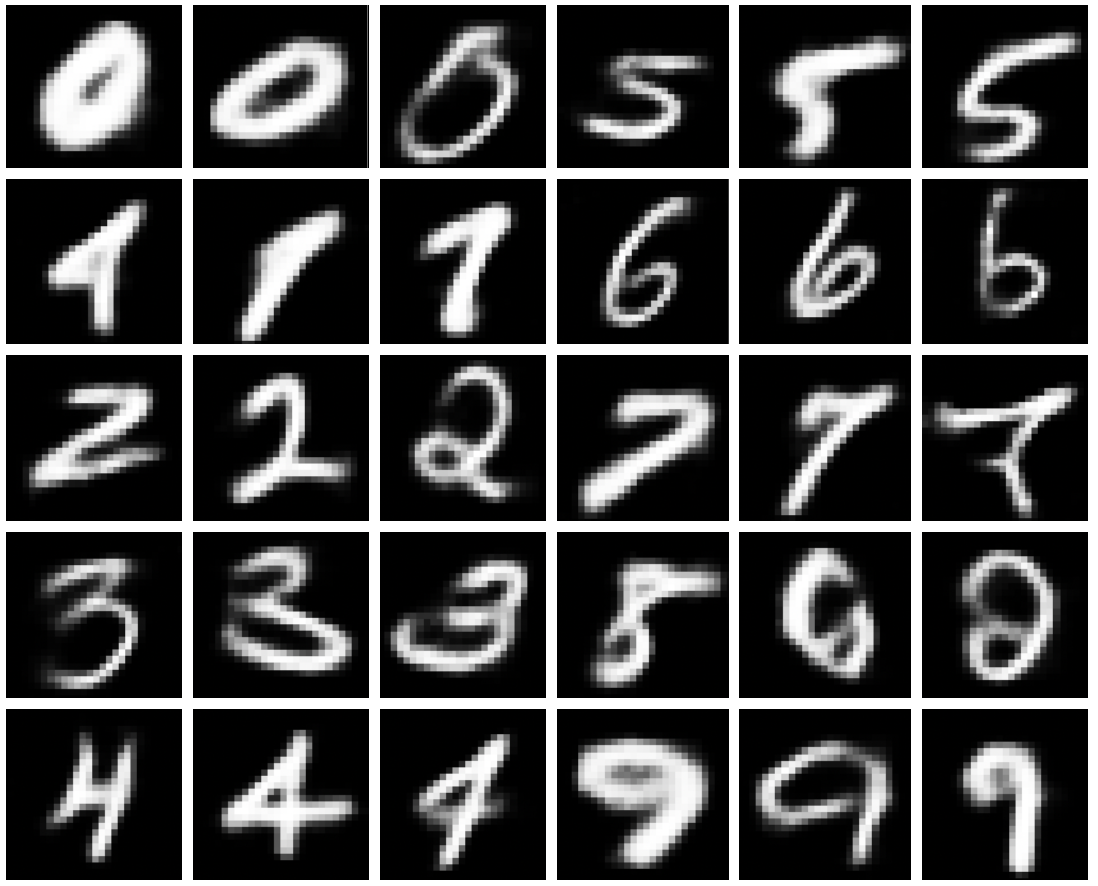}
\centering
\caption{Unconditioned latent model samples (ordered).}
\label{fig:unconditioned_samples}
\end{figure}

Samples generated with the unconditioned latent model are shown in figure \ref{fig:unconditioned_samples}.
Each image has been obtained by sampling from the quantum diffusion model and decoding each sample with a pre-trained decoder. The characteristics of the model used to generate these samples are reported in the second column table \ref{tab:params}. The value of the FID and WaM between the original and generated datasets can be found in the second column of table \ref{tab:metrics}.

\subsection{Conditioned latent model}
\begin{figure}[htb]
\includegraphics[width=0.443\textwidth]{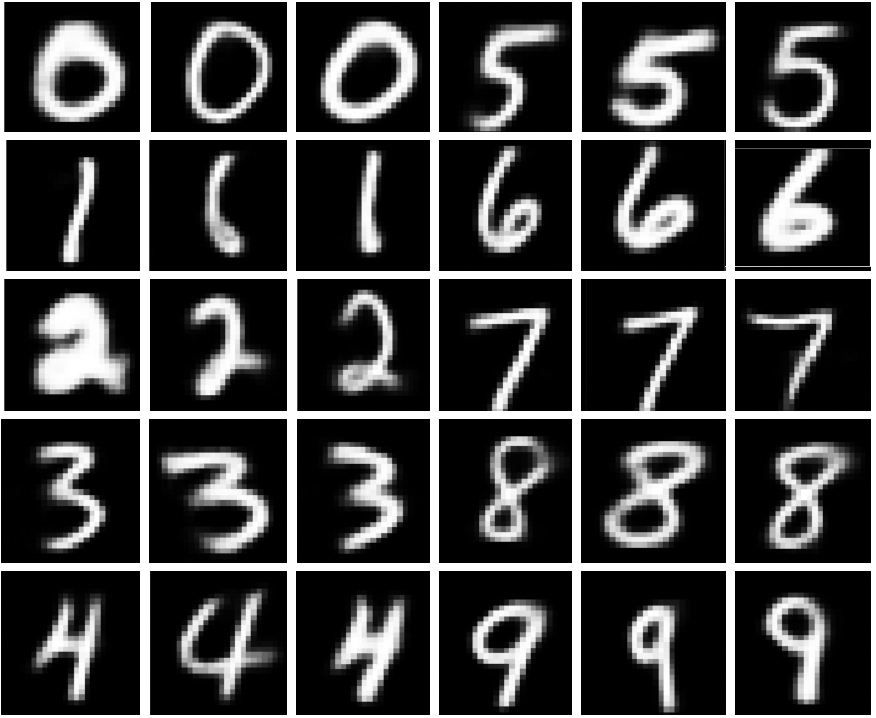}
\centering
\caption{Conditioned latent model samples.}
\label{fig:conditioned_samples}
\end{figure}

\begin{figure*}[h!]
\includegraphics[width=0.9\textwidth]{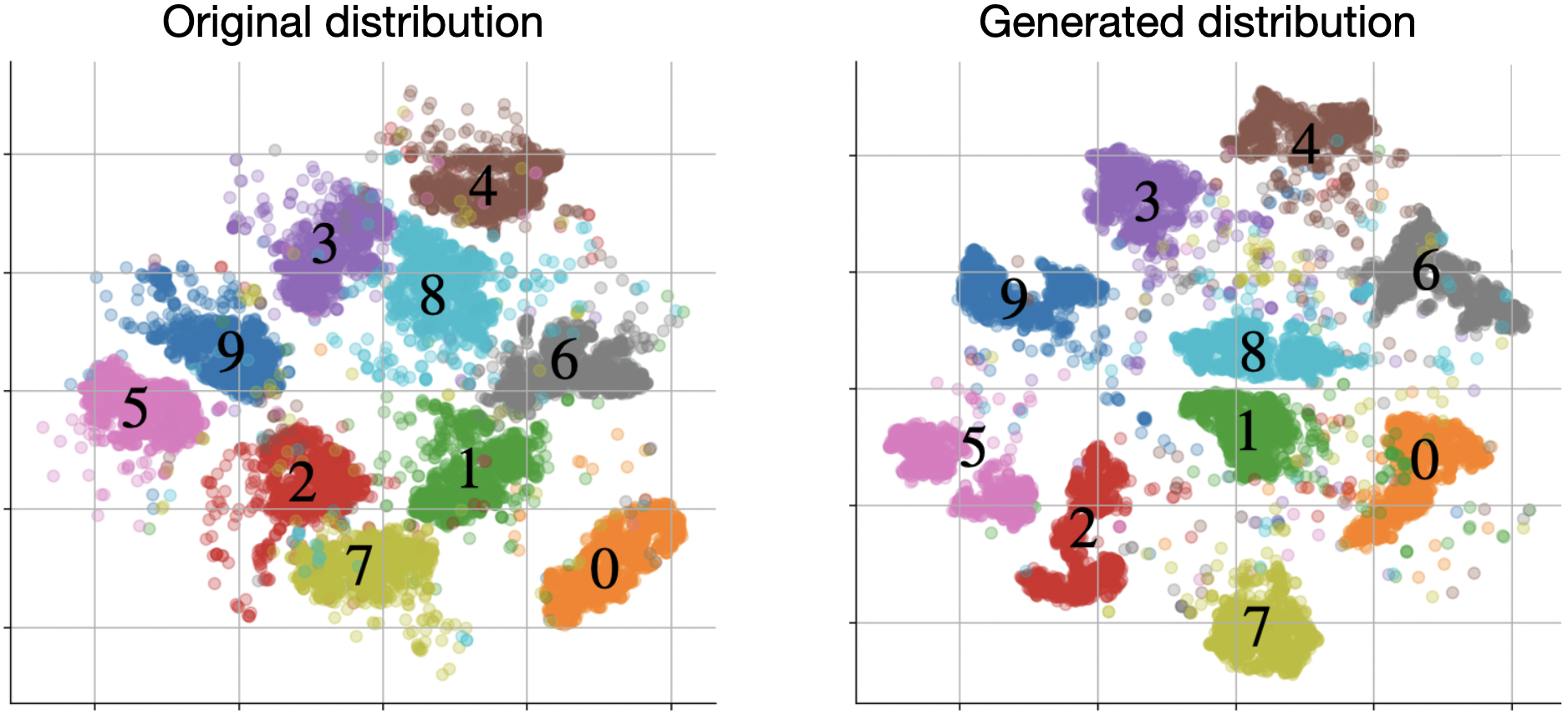}
\centering
\caption{2D representation (t-SNE) of the original distribution and generated distribution in the latent space (8D). Each digit is generated separately by conditioning the model.}
\label{fig:tsne}
\end{figure*}

Samples generated with the conditioned latent model are shown in figure \ref{fig:conditioned_samples}. Conditioning the model improves the samples quality. This improvement can be attributed to the fact that, in the conditioned model, the guiding process produces a clearer differentiation among the digits during sampling. Moreover, adding four qubits increases the dimension of the Hilbert space of the circuit and the number of trainable parameters, thus giving more expressive power to the PQC. The characteristics of the model used to generate these samples are reported in the last column of table \ref{tab:params}.
A comparison between the original and generated dataset in the latent space is reported in figure \ref{fig:tsne}, where we have used t-SNE to generate a 2D representation.
To assess conditioning, we use the ROC-AUC as detailed in section \ref{sec:evaluation}. The area under the curve (AUC) values for the ROC curves, specific to each digit, are presented in Table \ref{tab:AUC}. The digits ``one" and ``zero" exhibit the highest AUC values, which can be attributed to their simplicity and minimal overlap with other digits.
The value of the FID and WaM between the original and generated datasets can be found in the third column of table \ref{tab:metrics}. This is computed between the overall distributions, hence not considering the separation of digits. 
\begin{table}[ht]
\centering
\begin{tabular}{|c|c||c|c|}
\hline
Digit & AUC & Digit & AUC \\
\hline
0 & 0.951 & 5 & 0.899 \\
1 & 0.977 & 6 & 0.975 \\
2 & 0.983 & 7 & 0.928 \\
3 & 0.989 & 8 & 0.908 \\
4 & 0.857 & 9 & 0.933 \\
\hline
\end{tabular}
\caption{ROC-AUC values for each digit.}\label{tab:AUC}
\end{table}
\newpage
\section{Quantum hardware execution}\label{sec:hardware}
\begin{figure}[htb]
\includegraphics[width=0.49\textwidth]{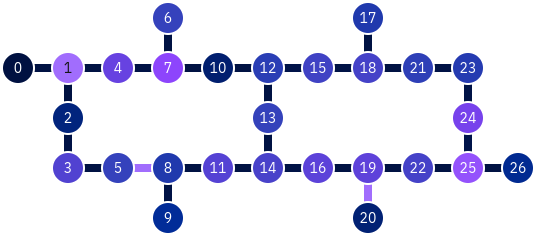}
\centering
\caption{Architecture of IBM\_hanoi quantum computer. The C-NOT connectivity is reported. The colors of single qubits and their connections represent respectively the $T1$ and C-NOT error probabilities. Darker colours represent a lower T1, in the range $[47, 265] \mu s$. For the C-NOT errors, darker colors represent a lower error rate, in the range  $[3.2 \times 10^{-3},1]$. Calibration data are referred to 19/10/2023.}
\label{fig:hanoi}
\end{figure}

Executing quantum circuits on current quantum devices poses significant challenges even when utilizing cutting-edge quantum hardware. The primary obstacles stem from the substantial error rates associated with quantum gates, particularly the C-NOT gates required for entangling qubits. 
A second problem comes from the short qubit relaxation time $T1$ \cite{carroll2022dynamics}. This value, also known as qubit lifetime, provides an estimate of how long a qubit can be used effectively in quantum computations before it becomes too prone to errors caused by decoherence. 
Another critical issue regards the quantum computer's architectural connectivity, since it's impossible to apply C-NOT gates between all qubit pairs. 
Contemporary NISQ devices provide connectivity exclusively between neighboring qubits, typically organized in linear or circular configurations. Figure \ref{fig:hanoi} illustrates the architecture of the quantum chip ibm\_hanoi\footnote{ https://quantum-computing.ibm.com/services/\\resources?system=ibm\_hanoi, 2021} employed in this study.
\\
When interactions between non-connected qubits are necessary, SWAP gates must be introduced in the circuit to exchange the quantum states of two qubits. It's worth noting that each SWAP gate requires three C-NOT gates.\\
Given these constraints, achieving significant results using the circuit described in section \ref{sec:QDM} is not possible. For the implementation of a QDM on quantum hardware, several modifications are essential to reduce the circuit's complexity.
\subsection{Quantum circuit adaptation}
In this section, we outline the modifications made to the quantum circuit to address the issues introduced before.
Regarding the connectivity, it's worth noting that the PQC ansatz proposed in Section \ref{sec:QDM} necessitates only interactions between neighboring qubits when they are arranged in a circular topology. However, in our quantum hardware configuration (as depicted in Figure \ref{fig:hanoi}), arranging six qubits in a circular pattern is not feasible. To mitigate this, we removed the last C-NOT gate from each entangling layer, thereby requiring interactions solely between neighboring qubits arranged linearly. This adjustment eliminated the need for SWAP gates within the circuit, significantly reducing the overall count of C-NOT gates.
\begin{figure*}[h!]
\includegraphics[width=\textwidth]{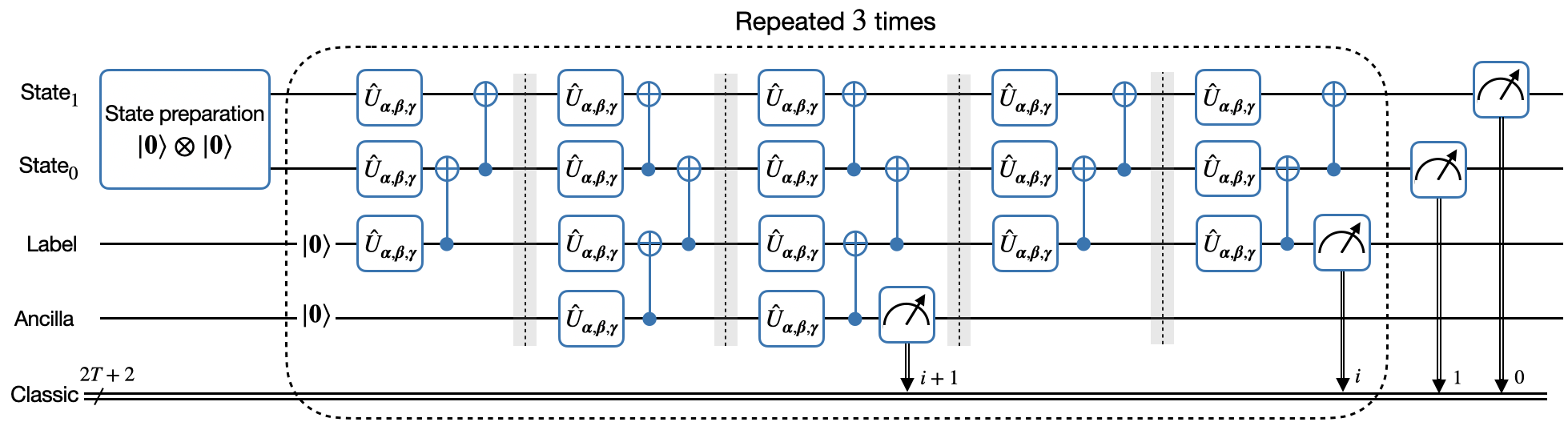}
\centering
\caption{Quantum hardware adapted PQC. The central part of the circuit is repeated three times in total. State preparation is necessary only to encode the initial noise. Similarly, the state qubits are measured only at the end.}
\label{fig:reduced_circ}
\end{figure*}\\
To reduce the circuit depth we have applied several changes. The first important change regards the number of denoising step. In fact, the complete quantum circuit is composed by a repetition of the denoising circuit for each denoising step. We found out that it is still possible to generate samples using just three denoising steps.\\
The second bigger change regards using a smaller denoising circuit, by reducing the number of layers. In particular we have used a reverse-bottleneck ansatz (section \ref{sec:QDM}), reported in figure \ref{fig:reduced_circ}. To restore the label qubit after its measurement, a conditional Pauli $X$ gate is applied.\\
To reduce the circuit depth further, we have also rearranged the C-NOT gates disposition in the entangling layers. In particular, we first apply all the C-NOT gates with an even control qubit and then the ones with an odd control qubit. This way only two circuit moments are required for each entangling layer.\\
The expressive power of the quantum circuit is highly reduced by all these adaptations, therefore, we have reduced the complexity of the task. We have restricted the problem to the conditioned generation of only zeros and ones, this way only one qubit is required to encode the label.
We have also reduced the latent space dimension from $8$ to $4$,  this way only two qubits are required to encode the state. The final circuit is thus composed of a total of four qubits and 37 C-NOT gates (36 for the PQC plus one for the amplitude encoding of the initial noise).

\subsection{Results}
\begin{figure*}[ht]
     \centering
     \includegraphics[width=0.9\textwidth]{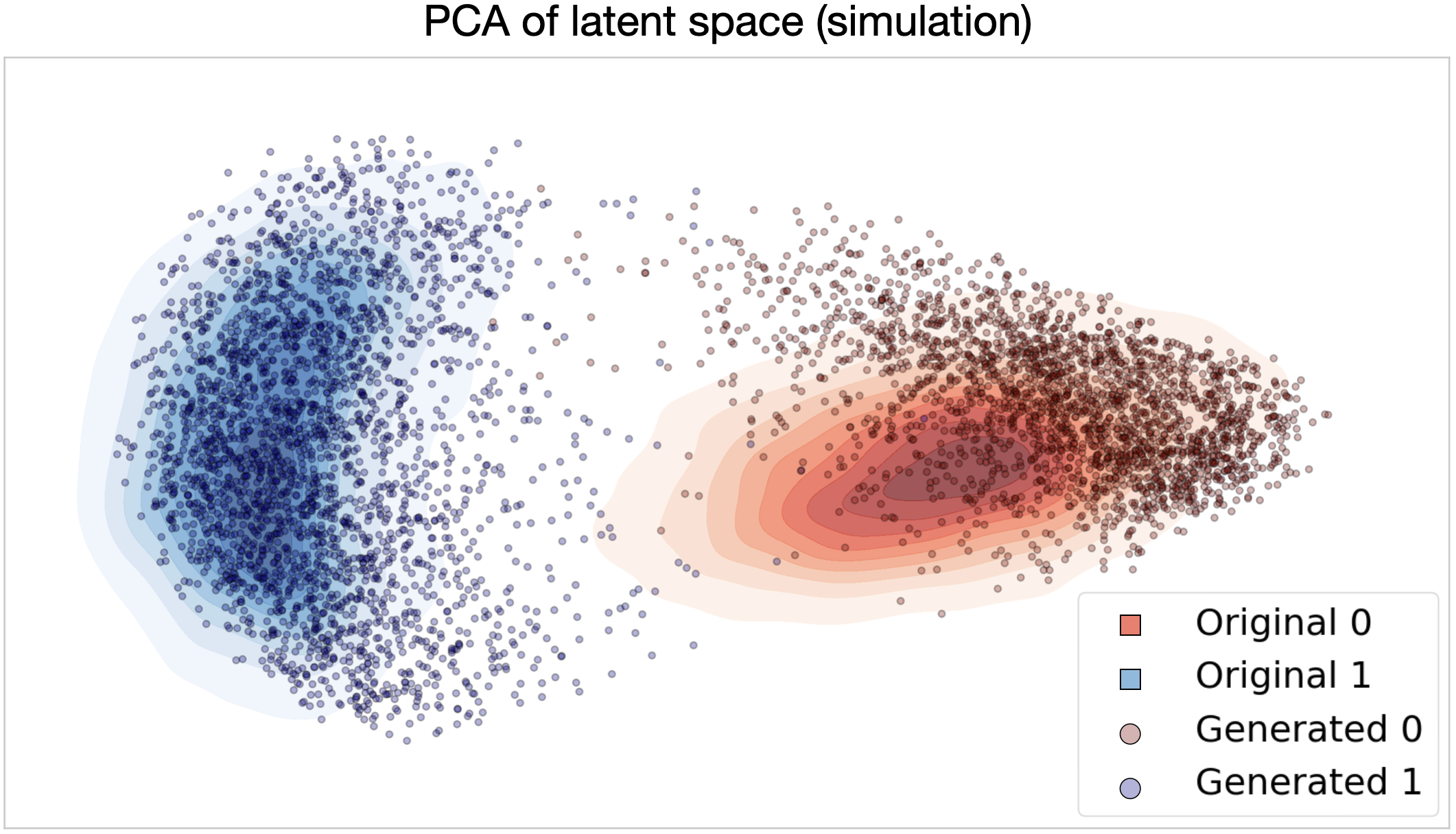}
     \caption{2D representation (PCA) of the latent vectors of the original distribution (continuous plots) and generated distribution with simulations (circles). The original distribution is composed by all the ``zeros" and ``ones" handwritten digits of the MNIST dataset. The generated distribution contains $5\times 10^3$ samples of each class.}
     \label{fig:pca_sim}
\end{figure*}

The hardware adapted PQC has been trained in simulation. Training on hardware devices is unfeasible on NISQ devices because of the high level of noise that makes the computation of gradients too imprecise. The training has been performed in the same way as described in section \ref{sec:conditioned} using ``zero" and ``one" handwritten digits.
The trained weights can be used for sampling on the real device. Before the hardware execution we have tested the trained algorithm in noiseless simulations. 
We have generated $5\times 10^3$ latent vectors of both ``zero" and ``one" digits in order to compare their distribution with respect to the latent vectors of the train set. 
Figure \ref{fig:pca_sim} shows the two dimensional PCA representation of the latent space for original and generated latent vectors. It is possible to observe that the model is able to correctly generate latent vectors with the distribution of the ``one" digits. On the other hand, for the ``zero" digits, there is not a full overlap between the two distributions. The model can generate samples that belong exclusively to a specific region within the original distribution, although the generated samples show a good variability. We have observed similar results in different training executions.

\begin{figure*}[ht]
     \centering
     \includegraphics[width=0.8\textwidth]{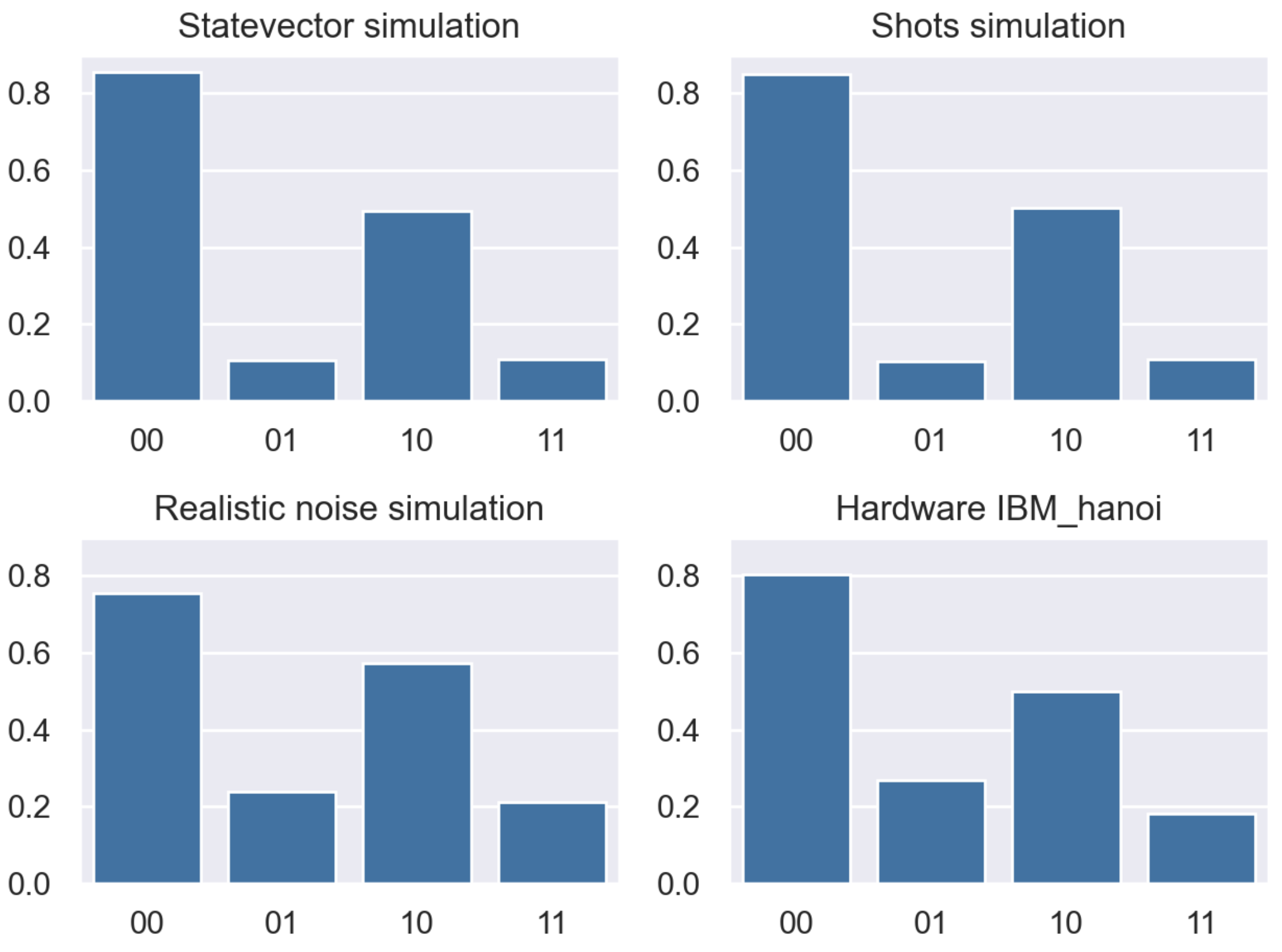}
     \caption{State amplitudes of a generated latent vector obtained under different noise conditions: state vector simulation, shots simulation, realistic noise simulation, quantum hardware.}
     \label{fig:shots}
\end{figure*}

\begin{figure*}[ht]
     \centering
     \includegraphics[width=0.6\textwidth]{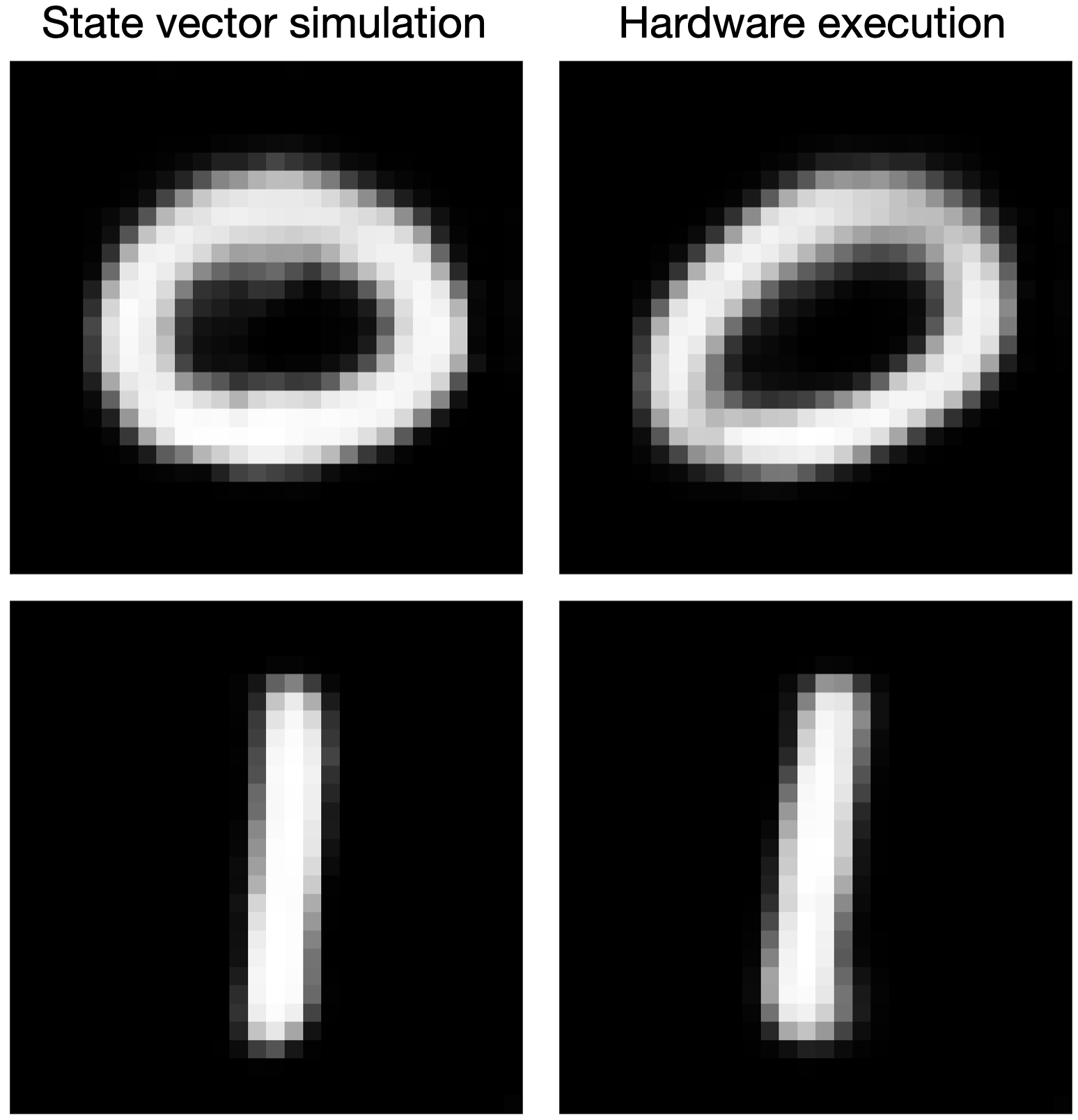}
     \caption{Generated images of a ``zero" and a ``one" handwritten digits. The latent vector has been generated with both a non noisy state vector simulation and with real quantum hardware in order to study the effects of noise.}
     \label{fig:images}
\end{figure*}

\begin{figure*}[ht]
     \centering
     \includegraphics[width=0.9\textwidth]{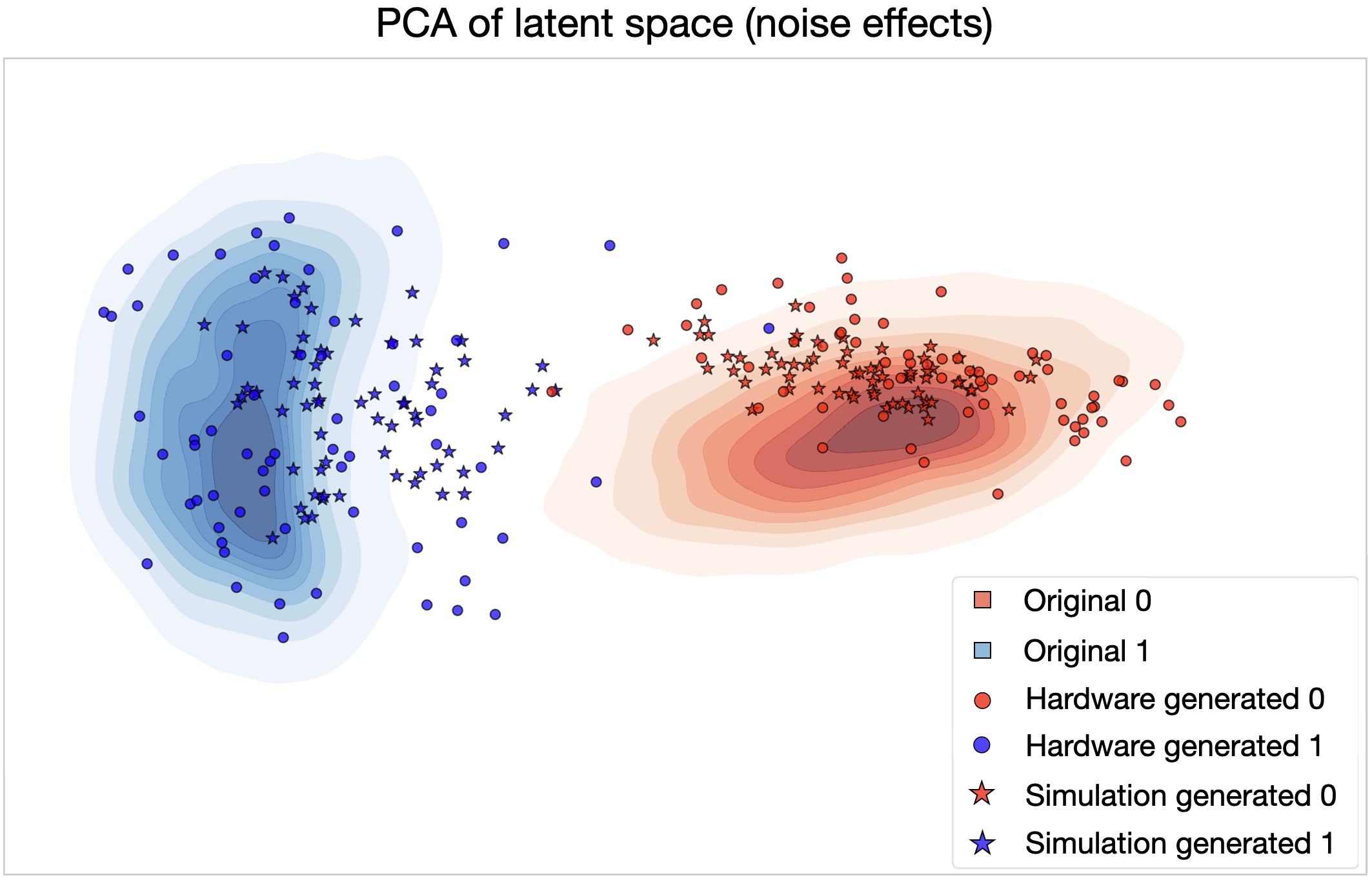}
     \caption{2D PCA representation of the latent vectors distribution obtained from the original MNIST dataset (continuous plots), generation with quantum hardware (stars) and generation with non-noisy shots simulation (circles). The generated distributions contain $64$ samples generated with $3.2\times 10^5$ shots and starting from the same initial noise.}
     \label{fig:pca_hard}
\end{figure*}

In order to study the effects of noise we have generated the latent vector of a ``zero" digit in four different ways: state vector simulation, shots simulation, realistic noise simulation and quantum hardware.
The realistic noise simulation has been performed with the Qiskit built-in noise model, that uses the calibration data to perform a shots simulation with realistic noise. 
Apart from the state vector simulation, all other methods require to measure the output states. The need to perform these measurements creates some difficulties, since the final state depends on the results of the measurement of the ancillary qubits. Six measurements are performed before the final state, three for the ancillary qubit and three for the label qubit. This means that, starting from the same initial noise, it is possible to obtain $64$ possible final states. In order to sample them all, it is necessary to execute the circuit over a large number of shots. In this study, we have performed $3.2\times 10^5$ shots for every circuit, obtaining, on average, $500$ shots for each possible final state. 
Figure \ref{fig:shots} reports the measurements frequency for each of the methods described before. It is possible to observe how the shot noise obtained with the shots simulation is negligible, with no significative difference with respect to the state vector simulation. The difference between the hardware execution and the state vector simulation is more relevant. The noise changes the amplitudes of the states $01$ and $11$ from about $0.1$ to $0.2$. This is an expected result, as decoherence, due to unwanted interactions with the environment, tends to bring the output state closer to the maximally mixed state. The realistic noise simulation is in good accordance with the real result, with the differences being in the range $5-10\%$.
\\
The effects of noise are relevant but do not influence the quality of the generated samples. Figure \ref{fig:images} shows the comparison of a ``zero" and a ``one" digit, generated with state vector simulation and on quantum hardware. The latent vectors have been decoded with the classical decoder. It is possible to observe how, in the presence of noise, the generated images change slightly, but still showing all the features of the correct class.
\\
We have used the results obtained on quantum hardware to study the effects of noise on the variability of the generated samples.
Figure \ref{fig:pca_hard} reports a two dimensional PCA representation of the latent space for the original and generated dataset, both in simulation and on quantum hardware. In both cases, the generated latent vectors have been obtained executing $3.2\times 10^5$ shots starting from the same initial noise. With this process, as previously explained, it is possible to obtain $64$ samples for each class.\\
In figure \ref{fig:pca_hard} it is possible to observe that noise reduces the variability of the generated samples. This is an opposite result with respect to the classical case, where introducing noise after each denoising step helps increasing the sample variability \cite{10081412}. The reduced variability with noise can be attributed to decoherence, which converges all final states towards the maximally mixed state.
Moreover, measurement errors tend to mix different samples, thus averaging them. This result highlights how the intrinsic differences between classical and quantum noise lead to differences in the behavior of classical and quantum generative algorithms.

\section{Conclusions}
In this paper, we introduce and analyze a novel quantum diffusion model, consisting of two distinct variants. The first is a comprehensive quantum model designed to generate classical or quantum data directly. The second is a latent classical-quantum model aimed at generating classical data. Both models can be conditioned using additional qubits to encode labels, thereby enhancing their practical utility. In our experiments we have demonstrated the practical applicability of these models on current NISQ devices.

The results obtained show that quantum diffusion models require fewer parameters with respect to their classical counterpart. For the latent model, this allows us to generate samples of similar quality to those generated by classical algorithms. Moreover, in the full quantum case, unlike classical models, it is possible to generate distributions whose features scale exponentially with the number of qubits. Although in simulation and on current NISQ devices we can represent a number of features comparable to the classical case, in the future the model might allow us to approximate probability distributions that are not classically tractable, including quantum datasets.
At this point, only the sampling part can be performed on quantum hardware. However, with the expectation of noiseless devices becoming available, it will be possible to train our algorithm in a fully quantum fashion, leveraging the parameter-shift rules.

A currently open problem remains to determine if it is possible to define a diffusion process directly in the Hilbert space, in order to implement a fully quantum pipeline that doesn't require classical computations.

\subsection*{Acknowledgements}
This work is partially supported by ICSC – Centro Nazionale di Ricerca in High Performance Computing, Big Data and Quantum Computing, funded by European Union – NextGenerationEU.

\subsection*{Data Availability Statement}
Both the datasets and the code used for this study are available on request by contacting the authors.

\clearpage
\onecolumn
\printbibliography
\end{document}